\documentclass[preprint,amsfonts,amsmath,amssymb]{revtex4}
\usepackage{graphicx}
\usepackage{dcolumn}
\usepackage{bm}

\begin{document}
\title{Thermal Casimir Effect in Kerr Space-time}
\author{Anwei Zhang}
\email{awzhang@sjtu.edu.cn}

\affiliation{%
Institute of Physics and Key Laboratory of Low Dimensional
Quantum Structures and Quantum Control of Ministry of Education,
Hunan Normal University, Changsha, Hunan 410081, China
}%

\begin{abstract}
  We investigate the thermal Casimir effect of a massless scalar field for two parallel plates moving in the equatorial orbit in Kerr space-time. Under the assumption that the typical cavity size is much smaller than the orbital radius, proposed by Sorge, we deduce the analytical expression of the renormalized free energy in this curved space-time.  We also get the analytical representation for the renormalized internal energy, and find that there is a thermal correction to the Casimir energy, which depends on the proper temperature and the proper geometrical parameters of the plates. The asymptotic behavior of the Casimir free energy,
entropy and internal energy at low temperature is also investigated.
\end{abstract}


\maketitle

\section{Introduction}

The Casimir effect \cite{Casimir}, discovered more than 60 years ago, is the
attraction between two neutral, perfectly conducting plates in vacuum.
In classical electrodynamics, the force should be equal to zero. However,
the quite remarkable result depends on the constant $\hbar$ . So, it is a
purely quantum effect, and results from the modification of the electromagnetic
vacuum by the boundaries.

The past few years have seen spectacular developments in Casimir effect, both theoretically
and experimentally \cite{bordag}. Some spaces with nontrivial topology, flat or curved, are also
new elements which have to be taken into account \cite{dewitt,ford,Al,sorge05}. A number of authors have
investigated the Casimir
effect in the presence of gravitational backgrounds \cite{Milton,Bimonte,Saharian,Elizalde}. In particular, the thermal
correction to the Casimir energy of a massless scalar field in cosmological models
have been investigated \cite{Bezerra11,Bezerra14}.

Recently, Bezerra et.al \cite{Bezerra142} calculate the vacuum energy for a massless scalar field
confined between two parallel ideal plates in Kerr space-time, in the weak field approximation.
In that paper, thermal correction to the Casimir energy is also considered, up to second order
in $M/R$. Relaxing the assumption of weak field approximation, the Casimir energy considered above
is generalized by Sorge \cite{sorge14}.

On the basis of Sorge$'$ work, we will take into account the temperature, then deduce the
renormalized free energy in the cavity placed in the same way as in \cite{sorge14}, and find out the thermal correction to the Casimir energy, which should depend on temperature.
We will also investigate the role of the background space-time on the thermal corrections and the asymptotic behavior of thermal Casimir quantities in the low  temperature limit.

The outline of this paper is as follows: In Sec. II, we briefly recall the main results of \cite{sorge14}.
In Sec. III, we deduce the renormalized free energy, entropy and internal energy as well as their behavior at low temperature.
Besides, the role of the background space-time on the thermal corrections is discussed. A brief conclusion is given in the last
section. Throughout the text,
the natural units: $\hbar=c=G=1$, will be used.

\section{An Overview of Sorge$'$ Work}

Now let us follow the approach developed by Sorge \cite{sorge14}, in which the vacuum energy of a scalar massless
field confined in a Casimir cavity moving in a circular equatorial orbit in the Kerr space-time geometry is
investigated. First, consider the Kerr space-time metric which can be given in the Boyer-Lindquist
coordinates $(t, r, \theta, \phi)$, namely
\begin{eqnarray} \label{1}
ds^2=\bigg(1-\frac{2Mr}{\Sigma}\bigg)dt^2+\frac{4Mar}{\Sigma}\sin^2\theta dtd\phi
\newline -\frac{\Sigma}{\Delta}dr^2-{\Sigma}d\theta^2-\frac{A\sin^2\theta}{\Sigma}d\phi^{2},
\end{eqnarray}
where $\Sigma=r^{2}+a^{2}\cos^{2}\theta,\; \Delta=r^{2}+a^{2}-2Mr, \; A=(r^{2}+a^{2})\Sigma+2Mra^{2}\sin^{2}\theta$.
$M$ and $a=J/M$ are the mass and the specific angular momentum of the rotating body.

It is assumed that the plates move in a circular equatorial orbit with angular velocity $\Omega={d\phi}/{dt}$
and are placed orthogonally to the equator line of the rotating source. In this case $(\theta=\pi/2)$, we have
$\Sigma=r^{2}$, \;$\Delta=r^{2}+a^{2}-2Mr$, \;$A=(r^{2}+a^{2})r^{2}+2Mra^{2}$.
The 4-velocity of a comoving observer can be expressed as $w^{\mu}=C(\Omega)(1, 0, 0, \Omega)$ in the Boyer-Lindquist
coordinates, where $C(\Omega)$ can be get from the normalization condition $g_{\mu\nu}w^{\mu}w^{\nu}=1$, namely
\begin{eqnarray}\label{2}
C(\Omega)=\bigg[\frac{r^{2}\Delta}{A}\bigg(1-\frac{A^{2}}{r^{4}\Delta}(\Omega-\omega_d)^{2}\bigg)\bigg]^{-1/2},
\end{eqnarray}
where $\omega_d=-g_{t\phi}/g_{\phi\phi}=2Mar/A$ is the dragging angular velocity of space-time around the rotating source.

Taking coordinate transformation $\phi'=\phi-\Omega t$, the metric becomes
\begin{eqnarray}\label{3}
g'_{tt}=C^{-2}(\Omega),\; g'_{t\phi}=g_{\phi\phi}(\Omega-\omega_d),\;
g'_{rr}=g_{rr},\; g'_{\phi\phi}=g_{\phi\phi}, \;g'_{\theta\theta}=g_{\theta\theta}.
\end{eqnarray}
It is worth noting that the 4-velocity of the observer is now $w'^{\mu}=C(\Omega)(1,0,0,0)$, and $g'=\textrm{det}(g'_{\mu\nu})=-r^{4}$. The inverse metric $g'^{\mu\nu}$ reads
\begin{eqnarray}\label{g}
  \bigg(\frac{\partial}{\partial s}\bigg)^{2} = \frac{A}{r^{2}\Delta}\bigg(\frac{\partial}{\partial t}\bigg)^{2} +2\frac{A}{r^{2}\Delta}(\omega_d-\Omega)(\frac{\partial}{\partial t})(\frac{\partial}{\partial \phi'})
  -\frac{1}{\Delta}C^{-2}(\Omega)\bigg(\frac{\partial}{\partial \phi'}\bigg)^{2}-\frac{\Delta}{r^{2}}\bigg(\frac{\partial}{\partial r}\bigg)^{2}-\frac{1}{r^{2}}\bigg(\frac{\partial}{\partial \theta}\bigg)^{2}.\nonumber \\
\end{eqnarray}

Define a Cartesian coordinates $(x, y, z)$ in the comoving frame and assume the $x$ axis tangent to the equator line
and the $y$ axis along the outward radial direction. Then in such a frame, in which the observer 4-velocity is the
same as $w'^{\mu}$,  metric ($\hat{g}_{\mu\nu}$) can be written as
\begin{eqnarray} \label{4}
ds^2=C^{-2}(\Omega)dt^2-\frac{A}{r^{3}}(\Omega-\omega_d)dtdx
\newline -\frac{A}{r^{4}}dx^2-\frac{r^{2}}{\Delta}dy^2-dz^2.
\end{eqnarray}
Note that there is a relation
\begin{equation}\label{r}
dx=rd\phi',\; dy=dr,\; dz=rd\theta.
\end{equation}

It is assumed that the typical size of the cavity is much smaller than the orbital radius, then the metric $\hat{g}_{\mu\nu}$ can
be considered as almost constant inside the cavity. Now the Klein-Gordon equation for the massless scalar field $\psi$ confined in the cavity
\begin{equation}\label{k}
 \frac{1}{\sqrt{-\hat{g}}}\partial_\mu[\sqrt{-\hat{g}}\hat{g}^{\mu\nu}\partial_\nu]\psi=0,
\end{equation}
can be written as
\begin{equation}\label{kgg}
 \hat{g}^{\mu\nu}\partial_{\mu}\partial_{\nu}\psi=0.
\end{equation}
Solving this equation with Dirichlet boundary
conditions imposed at the plates, we can get the eigenfrequencies of the scalar field
\begin{eqnarray}\label{5}
\omega_n=\frac{r}{\sqrt{\Delta} C^{2}(\Omega)}\bigg[\bigg(\frac{\pi n}{L}\bigg)^{2}+\frac{\Delta}{r^{2}}C^{2}
(\Omega)\bigg(\frac{\Delta}{r^{2}}k^{2}_y+k^{2}_z\bigg)\bigg]^{1/2},\nonumber \\
\end{eqnarray}
where $L$ is the coordinate separation between the plates, and $n$ is a natural number. After some calculations
(details see \cite{sorge14}), the renormalized
mean vacuum energy density can be obtained
\begin{eqnarray}\label{6}
\langle\varepsilon_{vac}\rangle_{ren}|_w=\varepsilon^{(0)}_{vac}\bigg[1-\frac{A^{2}}{r^{4}\Delta}(\Omega-\omega_d)\bigg]^{1/2},
\end{eqnarray}
where the subscript $w$ denotes the comoving oberver,
$\varepsilon^{(0)}_{vac}=-\frac{\pi^{2}}{1440L^{4}_p}$ is the vacuum Casimir energy density in flat space-time. Here,
\begin{equation}\label{7}
L_p=\int^{L}_0dx\bigg[-\hat{g}_{xx}+\frac{\hat{g}^{2}_{tx}}{\hat{g}_{tt}}\bigg]^{1/2}=\frac{L\sqrt{\Delta}}{r}C(\Omega)
\end{equation}
is the proper cavity length measured by the comoving observer.

\section{Thermal Casimir Effect}
\subsection{Renormalized Thermal Quantities}

In what follows we will take into account temperature in the cavity and derive the analytical representations for
the renormalized scalar Casimir free energy, entropy and internal energy. To begin with, we assume the
temperature field in the region limited by the two plates can be regarded as homogeneous and is independent of coordinate.
Since the thermal radiation temperature of Kerr black hole is only the function of the mass $M$ and the specific angular momentum $a$,
so this assumption can be easily satisfied. In the Matsubara formalism, it can be deduced that thermal correction to the free energy has the form \cite{geyer}
\begin{equation}
\Delta_T\mathcal{F}_0=k_BT\sum_J\ln\big(1-e^{-\beta\omega_J}\big),
\end{equation}
where $k_B$ is the Boltzmann constant, $J$ is a collective index.
In this paper, specifically, we write the above formula as
\begin{eqnarray}\label{8}
\Delta_T\mathcal{F}_0=\frac{1}{L}\sum^{\infty}_{n=1}\int^{\infty}_{-\infty}
\frac{dk_ydk_z}{(2\pi)^{2}}\int_V dxdydz\sqrt{\hat{g}_S}\cdot k_BT\ln\big(1
-e^{-\omega_n/k_BT}\big),
\end{eqnarray}
where $\hat{g}_S=-\det(\hat{g}_{\mu\nu})/\hat{g}_{tt}$. Note that the term $\frac{1}{L}\sum^{\infty}_{n=1}\int^{\infty}_{-\infty}
\frac{dk_ydk_z}{(2\pi)^{2}}\int_V dxdydz\sqrt{\hat{g}_S}$ is the phase volume and
in the case of $\hat{g}_S=1$, it becomes $S_0\sum^{\infty}_{n=1}\int^{\infty}_{-\infty}
\frac{dk_ydk_z}{(2\pi)^{2}}$ which is what we use in the calculation of thermal
correction in flat space-time. Here $S_0$ is the area of the plates.

\begin{equation} \label{9}
\int_V dxdydz\sqrt{\hat{g}_S}=S_0LC(\Omega)=V_p
\end{equation}
is the proper volume of the cavity measured by the comoving observer.

The total Casimir free energy can be written as
\begin{eqnarray}\label{10}
\mathcal{F}_0&=&E^{ren}_0+\Delta_T\mathcal{F}_0,
\end{eqnarray}
where
$E^{ren}_0=V_p\langle\varepsilon_{vac}\rangle_{ren}|_w$ is the renormalized
vacuum energy at zero temperature.

To calculate the thermal correction, it is convenient to introduce new variable $\tilde{k}_y, \tilde{k}_z$ and a parameter $\tilde{\beta}$
\begin{eqnarray}\label{11}
\tilde{k}_y=\frac{\Delta C(\Omega)}{r^{2}}k_y,\; \tilde{k}_z=\frac{\sqrt{\Delta}C(\Omega)}{r}k_z,\; \tilde{\beta}=\frac{1}{k_BT}\frac{r}{\sqrt{\Delta}C^{2}(\Omega)}.
\end{eqnarray}
Substituting Eqs.~(\ref{5}), (\ref{9}), (\ref{11}) in Eq.~(\ref{8}), we get
\begin{eqnarray}\label{12}
\Delta_T\mathcal{F}_0=\frac{S_0r^{4}}{2\pi\Delta^{2}C^{3}(\Omega)\tilde{\beta}}\sum^{\infty}_{n=1}
\int^{\infty}_{0}\tilde{k}_\perp d\tilde{k}_\perp\ln\bigg\{1
-e^{-\tilde{\beta}\big[(\frac{\pi n}{L})^{2}+\tilde{k}^{2}_\perp\big]^{1/2}}\bigg\},
\end{eqnarray}
where $\tilde{k}^{2}_\perp=\tilde{k}^{2}_y+\tilde{k}^{2}_z.$

Performing power series expansion of the logarithm in Eq.~(\ref{12}), we have
\begin{eqnarray}\label{13}
\Delta_T\mathcal{F}_0 = -\frac{S_0r^{4}}{2\pi\Delta^{2}C^{3}(\Omega)\tilde{\beta}}\sum^{\infty}_{n,m=1}\frac{1}{m}
\int^{\infty}_{0}\tilde{k}_\perp d\tilde{k}_\perp
e^{-m\tilde{\beta}\big[(\frac{\pi n}{L})^{2}+\tilde{k}^{2}_\perp\big]^{1/2}}.
\end{eqnarray}

By introducing the new variable
$z=\tilde{\beta}\bigg[\bigg(\frac{\pi n}{L}\bigg)^{2}+\tilde{k}^{2}_\perp\bigg]^{1/2},$
we get
\begin{eqnarray}\label{14}
\Delta_T\mathcal{F}_0 = -\frac{S_0r^{4}}{2\pi\Delta^{2}C^{3}(\Omega)\tilde{\beta}^{3}}\sum^{\infty}_{n,m=1}\frac{1}{m}
\int^{\infty}_{\frac{\pi n\tilde{\beta}}{L}}zdze^{-mz}
= -\frac{S_0r^{4}}{2\pi\Delta^{2}C^{3}(\Omega)\tilde{\beta}^{3}}\sum^{\infty}_{n,m=1}\frac{1+\frac{\pi mn\tilde{\beta}}{L}}{m^{3}}
e^{-\frac{\pi mn\tilde{\beta}}{L}}.\nonumber \\
\end{eqnarray}
Performing the summation in $n$, we can obtain,
\begin{eqnarray}\label{15}
  \Delta_T\mathcal{F}_0 &=&-\frac{S_0r^{4}}{16\pi\Delta^{2}C^{3}(\Omega)L^{3}}\sum^{\infty}_{m=1}\frac{(2\pi m\hat{\beta}+1)e^{2\pi m\hat{\beta}}-1}{(e^{2\pi m\hat{\beta}}-1)^{2}(m\hat{\beta})^{3}}.\nonumber \\
\end{eqnarray}
Here $\hat{\beta}=\tilde{\beta}/2L=1/\big(2L_Pk_BTC(\Omega)\big)$ has been introduced. It should be noted that
$S_0$ is mere coordinate area, the proper area is
\begin{equation}\label{16}
S_p=\int\int\sqrt{\hat{g}_{yy}\hat{g}_{zz}}dydz=\frac{r}{\sqrt{\Delta}}S_0.
\end{equation}
From Eqs.~(\ref{7}), (\ref{9}) and (\ref{16}), we have
\begin{equation}\label{17}
S_pL_p=S_0LC(\Omega)=VC(\Omega)=V_p,
\end{equation}
where $V$ is the coordinate volume of the cavity.
The Eq.~(\ref{15}) can be expressed in terms of the proper quantities and hyperbolic functions
\begin{eqnarray}\label{18}
\Delta_T\mathcal{F}_0=-\frac{S_p}{32\pi L^{3}_p}\sum^{\infty}_{m=1}\bigg[\frac{\coth(\pi m\hat{\beta})}{(m\hat{\beta})^{3}}
+\frac{\pi}{(m\hat{\beta})^{2}\sinh^{2}(\pi m\hat{\beta})}\bigg]+\frac{\zeta(3)S_p}{32\pi (L_p\hat{\beta})^{3}},
\end{eqnarray}
Where $\zeta(3)=\sum^{\infty}_{m=1}1/m^{3}$ is the Riemann zeta function.

Now let us consider
the asymptotic expression of Eq.~(\ref{18}) at high temperature (or large separation), namely, under the condition $\hat{\beta}\ll1$.
Performing Laurent series expansion of Eq.~(\ref{18}) in the deleted semi neighbourhood of $\hat{\beta}=0$, then taking
the summation in $m$ (we use zeta function regularization procedure to evaluate the infinite sum), thus we get
\begin{eqnarray}\label{19}
\Delta_T\mathcal{F}_0=-V_p\frac{\pi^{2}(k_BT)^{4}}{90}C^{4}(\Omega)+S_p\frac{\zeta(3)(k_BT)^{3}}{4\pi}C^{3}(\Omega)
-\frac{\pi^{2}S_p}{720L^{3}_p}.
\end{eqnarray}
Here $\zeta(4)=\pi^{4}/90$ has been used and $\zeta(-n)=-B_{n+1}/(n+1)$ vanishes at all even integers $n$.
Note that $B_{n+1}$ is a Bernoulli number.

Now we proceed with the renormalization of the Casimir free
energy at nonzero temperature. The
renormalization approach can go back to the paper \cite{geyer}, where the
renormalized  thermal correction  is
obtained by subtracting the terms $\alpha_0(k_BT)^{4}, \alpha_1(k_BT)^{3},$  and $\alpha_2(k_BT)^{2}$ which are contained in the asymptotic limit of high temperature of the free energy. Here the coefficients $\alpha_0, \alpha_1, \alpha_2 $ have
something to do with the geometrical parameters of the cavity and are expressed in terms of the heat kernel
coefficients \cite{bordag}. This approach is generalized to the case of the curved
space-time in Refs. \cite{Bezerra11, Bezerra14, Bezerra84}.

So, subtracting the first two terms in Eq.~(\ref{19}) from Eq.~(\ref{18}) we get the renormalized thermal correction
\begin{eqnarray}\label{21}
\Delta_T\mathcal{F}^{ren}=-\frac{S_p}{32\pi L^{3}_p}\sum^{\infty}_{m=1}\bigg[\frac{\coth(\pi m\hat{\beta})}{(m\hat{\beta})^{3}}
+\frac{\pi}{(m\hat{\beta})^{2}\sinh^{2}(\pi m\hat{\beta})}\bigg]+V_p\frac{\pi^{2}(k_BT)^{4}C^{4}(\Omega)}{90}.
\end{eqnarray}

In order to find the physical significance of the last term in Eq.~(\ref{21}), let us calculate the free energy density of
black-body radiation, expressed by
\begin{eqnarray}\label{22}
f_{bb}(T)=k_BT\int^{\infty}_{-\infty}\frac{dk_xdk_ydk_z}{(2\pi)^{3}}\ln(1-e^{-\omega_n/k_BT}).
\end{eqnarray}
Substituting Eqs.~(\ref{5}) and (\ref{11}) in Eq.~(\ref{22}) and performing the integration, we get
\begin{eqnarray}\label{23}
f_{bb}(T)=-\frac{\pi^{2}[k_BTC(\Omega)]^{4}}{90}.
\end{eqnarray}
So, the last term in Eq.~(\ref{21}) is just $-V_pf_{bb}(T)$.
Note that $T$ is the coordinate temperature, $C^{-1}(\Omega)$ is the factor of gravitational redshift
which can be get form the metric Eq.~(\ref{4}). So, $TC(\Omega)$ is actually the proper temperature $T_p$,
then the black-body free energy density can be written as
\begin{eqnarray}\label{24}
f_{bb}(T)=-\frac{\pi^{2}(k_BT_p)^{4}}{90},
\end{eqnarray}
which is same in the form with the one in flat space-time.

Then, the total renormalized Casimir free energy has the form
\begin{eqnarray}\label{25}
\mathcal{F}^{ren}=E^{ren}_0-\frac{S_p}{32\pi L^{3}_p}\sum^{\infty}_{m=1}\bigg[\frac{\coth(\pi m\hat{\beta})}{(m\hat{\beta})^{3}}
+\frac{\pi}{(m\hat{\beta})^{2}\sinh^{2}(\pi m\hat{\beta})}\bigg]-V_pf_{bb}(T).
\end{eqnarray}
The last two terms are the renormalized thermal correction $\Delta_T\mathcal{F}^{ren}$. And, $\hat{\beta}$ is actually the quantity $1/(2L_Pk_BT_p)$.

Now, let us calculate the renormalized Casimir entropy. Here we define it as
\begin{eqnarray} \label{28}
S^{ren}=-\frac{\partial\mathcal{F}^{ren}}{\partial T_p}.
\end{eqnarray}
After some calculations, we get
\begin{eqnarray}\label{29}
S^{ren}=\frac{3k_BS_p}{16\pi L^{2}_p}\bigg\{\sum^{\infty}_{m=1}
\bigg[\frac{\coth(\pi m\hat{\beta})}{m^{3}\hat{\beta}^{2}}
+\frac{\pi}{m^{2}\hat{\beta}\sinh^{2}(\pi m\hat{\beta})}
+\frac{2\pi^{2}\coth(\pi m\hat{\beta})}{3m\sinh^{2}(\pi m\hat{\beta})}\bigg]-\frac{4\pi^{3}}{135\hat{\beta}^{3}}\bigg\},
\end{eqnarray}
which plays an important role in thermodynamic tests of the approach to the calculation of the Casimir free energy.

Next, we consider the renormalized internal energy, which is defined similarly as
\begin{eqnarray}\label{30}
U^{ren}(T)=-T^{2}_p\frac{\partial}{\partial T_p}\bigg(\frac{\mathcal{F}^{ren}}{T_p}\bigg).
\end{eqnarray}
From Eqs.~(\ref{25}) and (\ref{30}), one obtains
\begin{eqnarray}\label{31}
U^{ren}(T)=E^{ren}_0+\frac{S_p}{16\pi L^{3}_p}\bigg \{\sum^{\infty}_{m=1}\bigg[\frac{\coth(\pi m\hat{\beta})}{(m\hat{\beta})^{3}}
+\frac{\pi}{(m\hat{\beta})^{2}\sinh^{2}(\pi m\hat{\beta})}+\frac{\pi^{2}\coth(\pi m\hat{\beta})}{m\hat{\beta}\sinh^{2}(\pi m\hat{\beta})}\bigg]
-\frac{\pi^{3}}{30\hat{\beta}^{4}}\bigg \}.\nonumber \\
\end{eqnarray}
The last term on the right of the above equation is just the thermal correction to the Casimir energy,
and it depends on the proper temperature and the proper geometrical parameters of the plates.

\subsection{Discussion}

In what follows, we will investigate the role of the background space-time on the thermal corrections.
Consider first the case that the rotating body is a black hole. Then according to the fact that the thermal radiation temperature of Kerr
black hole, which depends on $M$, $a$, is coordinate temperature, and the Tolman relation, we can easily find that the thermal corrections to thermal Casimir quantities measured by comoving observer will vary depending on the different orbits. If, on the other hand, the region limited by the two plates is
in equilibrium with a single reservoir which comoves with the Casimir apparatus, the proper temperature and then the thermal corrections measured by comoving observer will be invariant with respect to different orbits and background space-times. In this case, we can deduce the expressions of the thermal corrections in
flat space-time from Eqs.~(\ref{25}), (\ref{29}), (\ref{31}).

Note that the eigenfrequencies $\omega_n$ used in Eq.~(\ref{8}) are the ones obtained under the assumption that
the typical cavity size is much smaller than the orbital radius \cite{sorge14}, so this assumption is the premise for the validity of our calculation and conclusion above. Besides, it is worth noting that this assumption does not mean that the space-time in the region limited by the two plates is flat, because the eigenfrequencies of the scalar field contain the terms influenced by curved space-time.

Relaxing this assumption, we can solve the Klein-Gordon equation exactly by using the inverse metric (\ref{g}). The result is
\begin{eqnarray}\label{kg}
  g'^{\mu\nu}\partial_\mu\partial_\nu\psi+\frac{2}{r^{2}}(M-r)\partial_r\psi =0.
\end{eqnarray}
Then using the relation (\ref{r}) and the method given in \cite{sorge14}.
We obtain
\begin{eqnarray}
\omega'_n=\frac{r}{\sqrt{\Delta} C^{2}(\Omega)}\bigg[\bigg(\frac{\pi n}{L}\bigg)^{2}
+\frac{\Delta}{r^{2}}C^{2}
(\Omega)\bigg(\frac{\Delta}{r^{2}}k^{2}_y
+2i\alpha k_y+k^{2}_z\bigg)\bigg]^{1/2},
\end{eqnarray}
where $\alpha=(M-r)/r^{2}$ . When $|\alpha|\ll 1, \omega'_n\approx \omega_n$, we can have the same conclusion.
Actually in this condition, the Eq.~(\ref{kg}) will go back to the Eq.~(\ref{kgg}) from which the eigenfrequencies $\omega_n$ are obtained.
Giving the assumption $L\ll r$ which is equivalent to $rL/r^{2}\ll 1$, and the relation $ML/r^{2}\ll 1$ which can be deduced from the assumption \cite{sorge14}, we can find $|ML/r^{2}-rL/r^{2}|\ll 1$ and then $|\alpha|\ll 1$. So our assumption and the condition $|\alpha|\ll 1$ are compatible.

\subsection{The Limit of Low Temperature}

Here, we consider the asymptotic behavior of the Casimir free energy, entropy and internal energy in low temperature (or small separation) limit which is equivalent to $\hat{\beta}\gg1$. In such limit, we have
\begin{equation}\label{33}
\coth(\pi m\hat{\beta})\approx1,\; \sinh(\pi m\hat{\beta})\rightarrow\infty.
\end{equation}
Thus, from Eq.~(\ref{25}) one obtains
\begin{equation}\label{34}
\mathcal{F}^{ren}\approx E^{ren}_0-\frac{\zeta(3)S_p(k_BT_p)^{3}}{4\pi}+V_p\frac{\pi^{2}(k_BT_p)^{4}}{90}.
\end{equation}
All corrections to Eq.~(\ref{34}) are exponentially small. This can be seen when we replace
the hyperbolic functions in Eq.~(\ref{25}) with exponential functions
and take $1/(e^{2\pi m \hat{\beta}}-1)$ as $e^{-2\pi m \hat{\beta}}$.
The leading exponentially small correction to be added to Eq.~(\ref{34}) can be obtained when taking $m=1$, namely
\begin{equation}\label{35}
-\frac{S_p}{2 L_p}(k_BT_p)^{2}e^{-\pi/(L_pk_BT_p)}.
\end{equation}

In a similar way, the low temperature behavior of the Casimir entropy is given by
\begin{equation}\label{36}
S^{ren}\approx\frac{3\zeta(3)}{4\pi}S_pk^{3}_BT^{2}_p-\frac{2\pi^{2}}{45}V_pk^{4}_BT^{3}_p.
\end{equation}
The leading exponentially small correction to be added has the form
\begin{equation}\label{37}
\frac{\pi k_B S_p }{2 L_p^{2}}e^{-\pi/(L_pk_BT_p)}.
\end{equation}
As can be seen from Eqs.~(\ref{36}) and (\ref{37}), the Casimir entropy goes to zero when
the temperature vanishes. So the third law of thermodynamics is satisfied.

For the Casimir inter energy at low temperature, one obtains
\begin{equation}\label{38}
U^{ren}(T)\approx\ E^{ren}_0+\frac{\zeta(3)S_p(k_BT_p)^{3}}{2\pi}-V_p\frac{\pi^{2}(k_BT_p)^{4}}{30}.
\end{equation}
The leading correction to Eq.~(\ref{38}) is given by
\begin{equation}\label{39}
\frac{\pi S_p(k_BT_p)}{2L^{2}_p}e^{-\pi/(L_pk_BT_p)}
\end{equation}
From Eq.~(\ref{38}) and (\ref{39}) we can find the Casimir internal energy goes to $E^{ren}_0$ when
the temperature vanishes.

\section{Conclusions}
In this paper, we have investigated the thermal Casimir effect of a massless scalar field confined in two
parallel ideal plates moving in the equatorial orbit in the Kerr space-time.
Under the assumption that the typical cavity size is much smaller than the orbital radius proposed by Sorge, we have deduced the analytical representation for the renormalized free energy. We calculate the free energy density of black-body radiation in this space-time and find that it has the same form with the one in flat space-time.
We also get the analytical expressions of the renormalized entropy and internal energy and find that in the presence of thermal bath, there is a modification in the Casimir energy, which depends on the proper temperature and the proper geometrical parameters of the plates.

Besides, we investigate the asymptotic behavior of the Casimir
free energy, entropy and internal energy in the limit of low  temperature. In particular, we find that when the temperature trends to zero, the Casimir entropy moves towards zero which is consistent to
the third law of thermodynamics, and the internal energy coincids with the zero-temperature Casimir energy.

To keep our calculation as clear and transparent as possible, we only consider the simplest case of massless scalar field.
The case of high spins can be done similarly. The paper \cite{Bezerra84}, where
the thermal Casimir effect is studied for neutrino and electromagnetic
fields in the closed Friedmann Universe, is pertinent.

 \begin{acknowledgments}
We would like to thank H. Yu,  W. Zhou and F. Sorge for a valuable discussion.
 \end{acknowledgments}

\end{document}